\documentclass[twocolumn]{aastex6}
\usepackage{multirow}
\usepackage{amsbsy}
\usepackage{wasysym}
\usepackage{newfloat}

\DeclareFloatingEnvironment[name={Supplementary Figure}]{suppfigure}

\begin{document}

\title{Gaia-DR2 confirms VLBA parallaxes in Ophiuchus, Serpens and Aquila}

\author{Gisela N.\ Ortiz-Le\'on\altaffilmark{1},  Laurent Loinard\altaffilmark{2,3}, Sergio A.\ Dzib\altaffilmark{1}, Marina Kounkel\altaffilmark{4}, Phillip A.\ B.\ Galli\altaffilmark{5}, John J.\ Tobin\altaffilmark{6,7}, Neal J.\ Evans II\altaffilmark{8}, Lee Hartmann\altaffilmark{9}, Luis F.\ Rodr\'{\i}guez\altaffilmark{2}, Cesar Brice\~no\altaffilmark{10}, Rosa M.\ Torres\altaffilmark{11} and Amy J.\ Mioduszewski\altaffilmark{12}
}

\email{gortiz@mpifr-bonn.mpg.de}

\altaffiltext{1}{Max Planck Institut f\"ur Radioastronomie, Auf dem H\"ugel 69,  D-53121 Bonn, Germany}
\altaffiltext{2}{Instituto de Radioastronom\'ia y Astrof\'isica,  Universidad Nacional Aut\'onoma de Mexico, Morelia 58089, Mexico }
\altaffiltext{3}{Instituto de Astronom\'ia, Universidad Nacional Aut\'onoma de M\'exico, Apartado Postal 70-264, 04510 Ciudad de M\'exico, M\'exico}
\altaffiltext{4}{Department of Physics and Astronomy, Western Washington University, 516 High St, Bellingham, WA 98225, USA}
\altaffiltext{5}{Laboratoire d'astrophysique de Bordeaux, Univ. Bordeaux, CNRS, B18N, all\'ee Geoffroy Saint-Hillaire, 33615 Pessac, France.}
\altaffiltext{6}{Homer L. Dodge Department of Physics and Astronomy, University of Oklahoma,  440 W. Brooks Street, Norman, OK 73019, USA }
\altaffiltext{7}{Leiden Observatory, PO Box 9513, NL-2300 RA, Leiden, The Netherlands}
\altaffiltext{8}{Department of Astronomy, The University of Texas at Austin, 2515 Speedway, Stop C1400, Austin, TX 78712-1205, USA }
\altaffiltext{9}{Department of Astronomy, University of Michigan, 500 Church Street,  Ann Arbor, MI 48105,  USA}
\altaffiltext{10}{Cerro Tololo Interamerican Observatory, Casilla 603, La Serena, Chile}
\altaffiltext{11}{Centro Universitario de Tonal\'a, Universidad de Guadalajara, Avenida Nuevo Perif\'erico No. 555, Ejido San Jos\'e Tatepozco, C.P. 48525, Tonal\'a, Jalisco, M\'exico.}
\altaffiltext{12}{National Radio Astronomy Observatory, Domenici Science Operations Center, 1003 Lopezville Road, Socorro, NM 87801, USA}

\begin{abstract}

We present Gaia-DR2 astrometry of a sample of YSO candidates in Ophiuchus, Serpens Main and Serpens South/W40 in the Aquila Rift, which had been mainly identified by their infrared excess with {\it Spitzer}. We compare the Gaia-DR2 parallaxes against published and new parallaxes obtained from our Very Long Baseline Array (VLBA) program GOBELINS.
We obtain consistent results between Gaia and the VLBA for the mean parallaxes in each of the regions analyzed here. We see small offsets, when comparing mean values, of a few tens of micro-arcseconds in the parallaxes, which are either introduced by the Gaia zero-point error or due to a selection effect by Gaia toward the brightest, less obscured stars. Gaia-DR2 data alone  conclusively places Serpens Main and Serpens South at the  same distance, as we first inferred from VLBA data alone in a previous publication. Thus, Serpens Main, Serpens South and W40 are all part of the same complex of molecular clouds, located at a mean distance of $436\pm9$~pc.  In  Ophiuchus,  both Gaia and VLBA suggest a small parallax gradient across the cloud, and the distance changes from $144.2\pm1.3$~pc to $138.4\pm2.6$~pc when going from L1689 to L1688. 

 \end{abstract}


\section{Introduction}\label{sec:intro}
     Observations of nearby regions rich in extremely young objects (for example Ophiuchus, Perseus, Taurus, Orion, Serpens, etc.) provide a wealth of information and clues to understand the paradigm for star formation. 
      In this context,  the knowledge of the distance to star-forming regions is essential for properly interpreting the observed properties of young objects, from pre-stellar cores to disks, as well as their environment. An advance in constraining the distance of these star-forming regions has been made with astrometric observations using Very Long Baseline Interferometry \citep[VLBI; see for example][]{Reid_2014}. 
In particular, our team has conducted a major program to measure individual distances during the last few years, via the trigonometric parallax, and proper motions to almost a hundred young stars in such nearby star-forming regions (the Gould's
Belt Distances Survey -- GOBELINS; \citealt{Loinard_2013}). The distance to young stars in Ophichus \citep{Ortiz2017Oph}, Serpens \citep{Ortiz2017Ser}, Orion \citep{Kounkel_2017}, Taurus \citep{Galli2018} and Perseus \citep{Ortiz2018}  is now known with $0.2 - 3\%$ accuracy, yielding relevant information about their three-dimensional  structure.  

On 2018, April 25 the Gaia mission published its second Data Release \citep[DR2;][]{Lindegren_2018,gaia_2018}. This release contains astrometric solutions for more than 1.3 billion stars with parallax uncertainties of $\sim0.7$~mas for a magnitude $G = 20$.  Several young stars within the star-forming regions studied by our program have parallax and proper motions measurements available in the DR2 catalog. Therefore, we can now compare the VLBI astrometry against Gaia, to validate the accuracy of our measurements. 

Here we analyze Gaia-DR2 and VLBA astrometry in Ophiuchus, Serpens and the Aquila Rift. Ophiuchus consists of a main cloud known as Lynds 1688  (L1688; see e.g.\ \citealt{Wilking_2008} for a recent review) and several filamentary clouds to the northeast (L1709 and L1704) and the southeast (L1689, L1712 and L1729). 
      The Serpens Molecular cloud and the Aquila Rift are  two cloud complexes projected close to each other in the plane of the sky.  The well studied Serpens Main cluster is embedded within the Serpens Molecular cloud \citep{Eiroa_2008}, while W40 \citep{Smith_1985} and the extremely young Serpens South cluster \citep{Gutermuth_2008} are embedded within the Aquila Rift, about 3$^{\rm o}$ to the south of Serpens Main. Because Serpens South has the most star-formation activity within the Aquila, it is often referred to as the Aquila or the Serpens-Aquila region (e.g.\ within the {\it Herschel} Gould Belt Survey program; \citealt[][]{Andre_2010}).
     
\section{VLBA observations}\label{sec:obs}

New Very Long Baseline Array (VLBA) observations toward Ophiuchus, Serpens Main and W40 (within the Aquila Rift) were performed in the period from 2016, August to 2018, October under project code BL175.
These data were taken at 5~GHz following the observing strategy described in detail in \citet[][]{Ortiz2017Oph} and \citet[][hereafter Papers I and II, respectively]{Ortiz2017Ser}.

The resulting astrometric parameters from the updated fits, including new sources, are given in Table \ref{tab:prlx-oph}. In addition to the 24 objects published in  Papers I and II, here we present new parallaxes for 6 objects. Following the approach outlined in \cite{Loinard2007}, the data are fitted with a model that assumes a uniform proper motion and has the following free parameters: parallax ($\varpi$), proper motions $(\mu_\alpha,\mu_\delta)$ and position at the reference epoch $(\alpha_0,\delta_0)$. The fits to sources that are members of wide binary systems also include acceleration terms, $(a_\alpha,a_\delta)$. 
We also update the orbital fits of our binary systems following \cite{Kounkel_2017}, however, in this paper we will  focus solely on parallax, and leave the discussion on the orbital parameters for a forthcoming publication. 

For Ophiuchus, the weighted mean of individual parallaxes in each cloud gives $\varpi=7.23\pm 0.14$~mas and  $\varpi=6.93\pm 0.06$~mas, for L1688 and L1689, respectively. Herein, and in the rest of the paper, the quoted errors on weighted mean parallaxes correspond to the standard deviation,  unless otherwise noted. These measurements suggest that, although small, there is a parallax gradient across this cloud, but the difference in parallax is only significant at $2\sigma$.

In Serpens/Aquila, we found $\varpi=2.31 \pm 0.01$~mas and $\varpi=2.23 \pm 0.10$~mas, for the Serpens Main and W40 clusters, respectively. In this case, the mean parallaxes are consistent between them within $1\sigma$. The average of all sources in both regions gives $\varpi= 2.30 \pm 0.05$~mas.

\section{Analysis of Gaia DR2 data}

We use the list of young stellar object (YSO) candidates derived with {\it Spitzer} by \cite{Dunham_2015} for Ophiuchus (292 sources), Serpens (227 sources) and Aquila (1319 sources). However, this sample suffers from contamination by background AGB stars \citep[between 25\% and 90\%;][]{Dunham_2015}, so their membership and nature as true YSOs must be confirmed. We also use the list of 316 YSOs compiled by \cite{Wilking_2008} for Ophichus, which are all 2MASS sources within the L1688 cloud, with detection in the $K_S$ band and other signs of youth, such as H$_\alpha$ or X-ray emission. The combined sample has 565 YSO candidates in Ophiuchus.

\begin{figure*}[!bht]
 {\includegraphics[width=0.34\textwidth,angle=0]{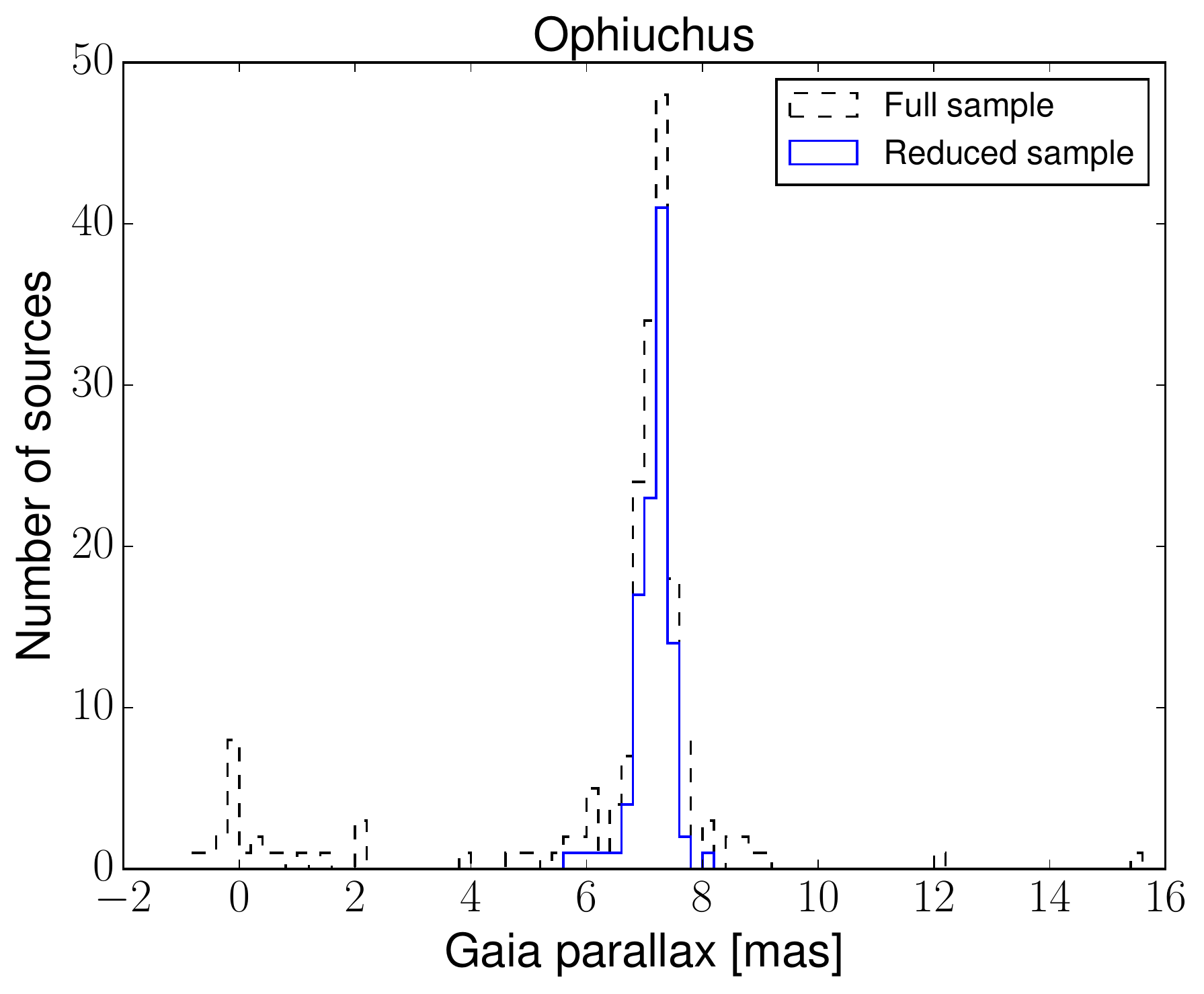}}
 {\includegraphics[width=0.34\textwidth,angle=0]{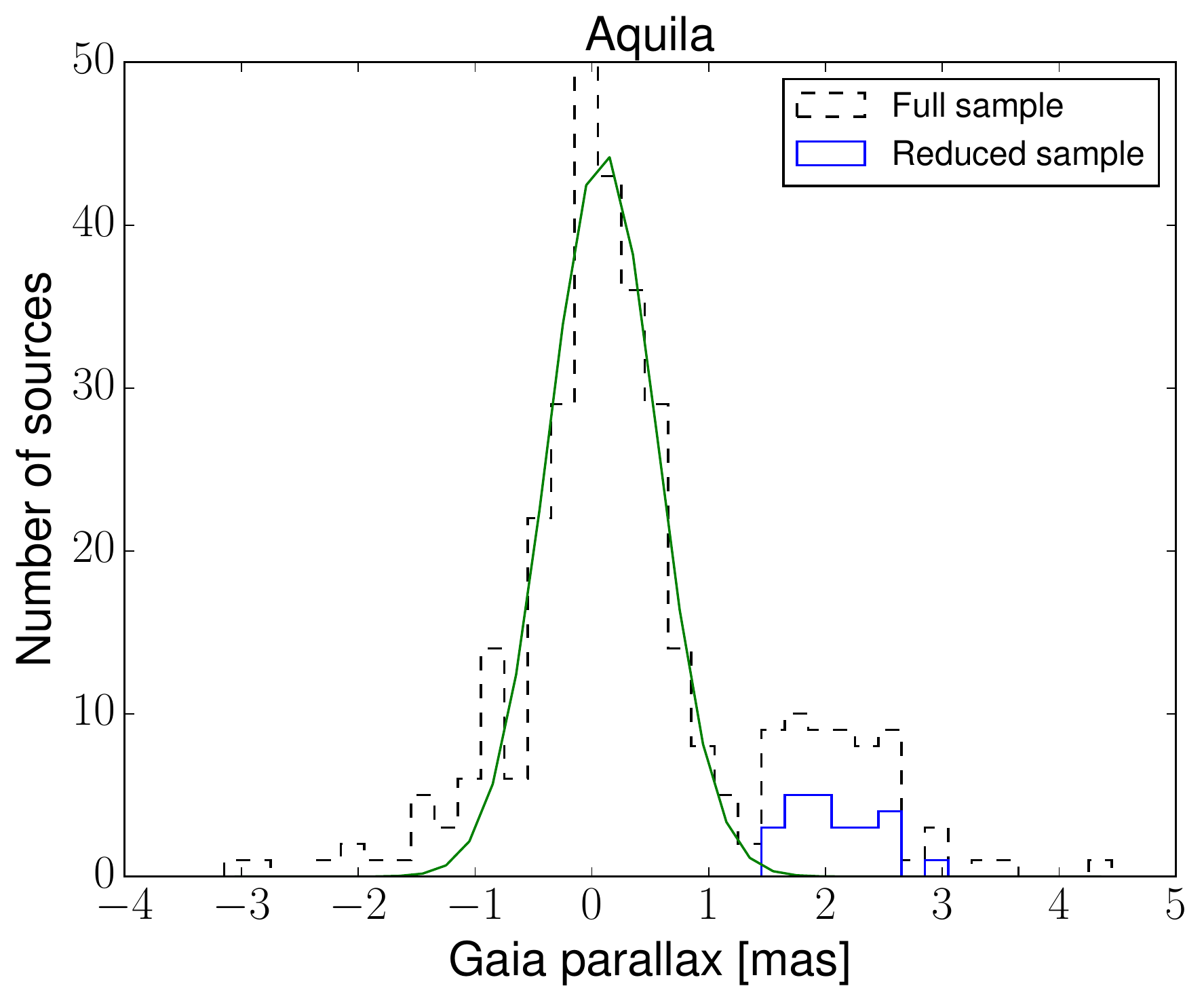}}
 {\includegraphics[width=0.34\textwidth,angle=0]{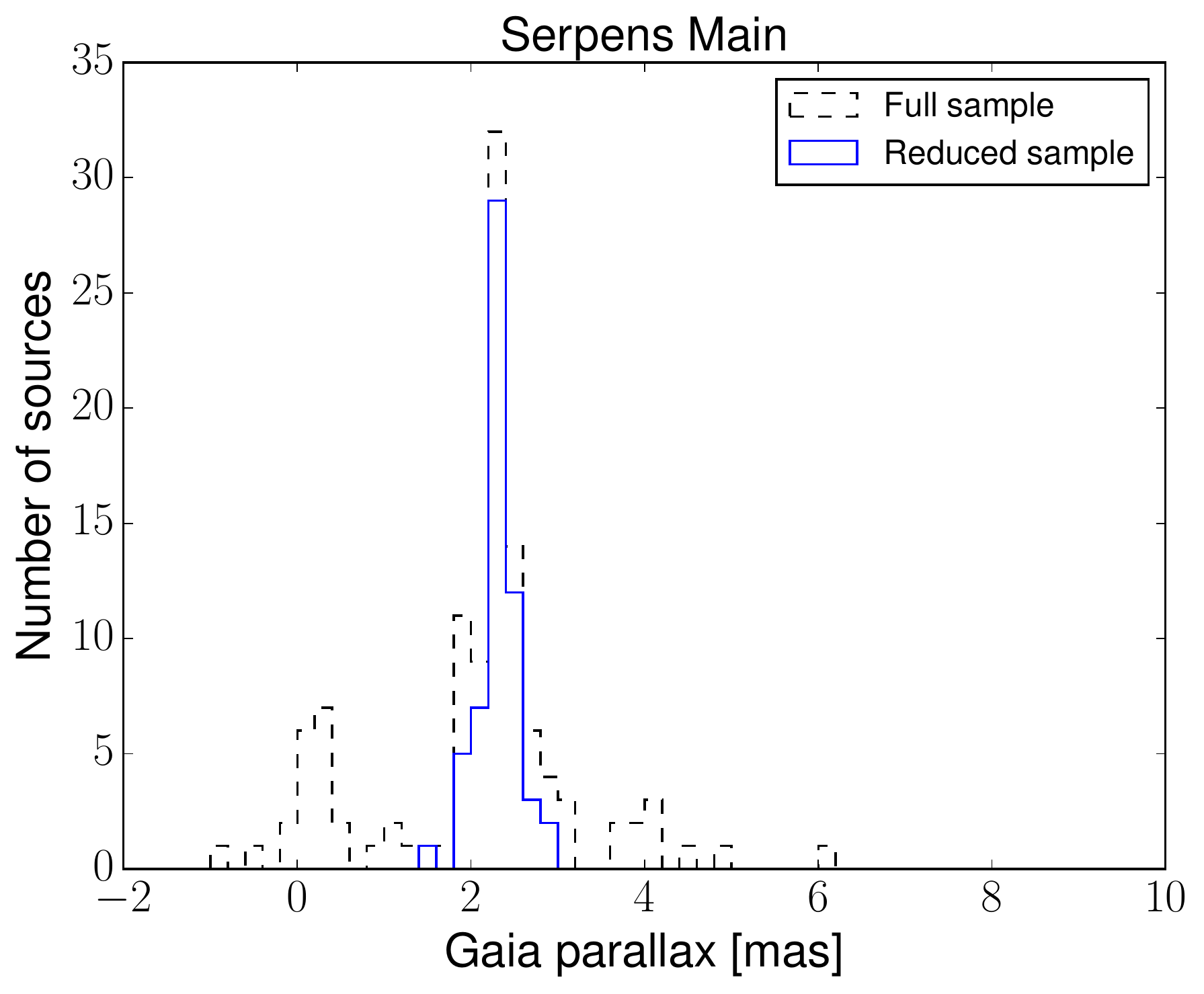}}
\caption{Gaia parallaxes from DR2 in Ophiuchus (left), Aquila (middle) and Serpens Main (right). The blue histograms show the distribution of parallaxes after cutting stars according to the criteria given in the text. The green line in the middle panel is a Gaussian fit to the parallax distribution with $\varpi\apprle1.4$~mas.}
\label{fig:gaia-prlx}
\end{figure*}

\begin{figure*}[!bht]
\begin{center}
 {\includegraphics[width=0.8\textwidth,angle=0]{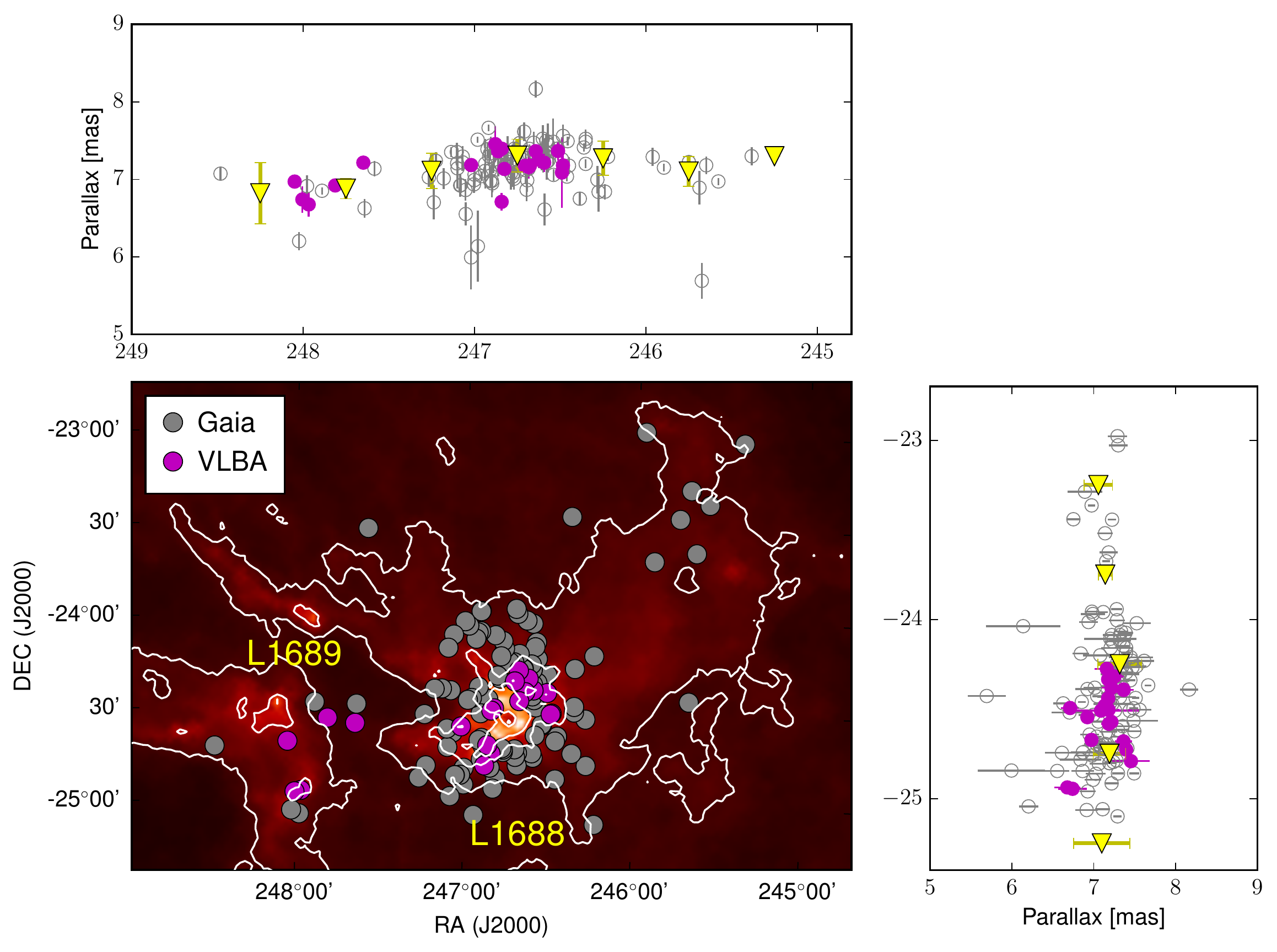}}
 \end{center}
\caption{Gaia-DR2 and VLBA parallaxes measured in Ophiuchus. The inner panel shows the spatial distribution of stars with
measured parallaxes, while the top and right panels show the parallaxes as a function of R.A.\ and declination, respectively. The grey open circles are stars with Gaia parallaxes, while the magenta filled circles are stars with VLBA parallaxes. The yellow triangles are the weighted mean of Gaia parallaxes determined for bins with a width of $0.5^{\rm{o}}$. The map in the background is an extinction map obtained as part of the COMPLETE project \citep{Ridge_2006}. The white contours indicate $A_V=$~4, 12 and 20.}
\label{fig:oph-plx-vs-pos}
\end{figure*}

We cross-matched the positions of the YSO candidates against Gaia DR2 positions using a search radius of $1"$ and the Virtual Observatory  tool TOPCAT \citep{Taylor2005}. In Ophiuchus, 191 sources of the {\it Spitzer}+2MASS catalog have astrometric solutions.  The distribution of Gaia parallaxes is shown in Figure \ref{fig:gaia-prlx} (left panel) and their positions on the plane of the sky in Figure \ref{fig:oph-plx-vs-pos}. We note  that this sample contains background and foreground contaminants with parallaxes $\apprle 3$ and $\apprge11$~mas (i.e.\ well outside the range from 120 to 150~pc reported in the literature as the distance to Ophiuchus). To remove these stars, we look at the distributions of proper motions and, following \cite{Ortiz2018} and \cite{Dzib2018b}, we cut all stars with proper motions outside $3\sigma$ from the mean, as determined by fitting a Gaussian model to the measured proper motions in each direction. Then, we produce a sample with reliable parallaxes by applying the recommended cuts given in \cite{Lindegren_2018}, which are designed to remove sources with poor or spurious astrometric solutions. Specifically, we applied the criteria expressed in equations C.1 and C.2 in appendix C of \cite{Lindegren_2018}.  After cutting also sources that lie well outside the region shown in Figure \ref{fig:oph-plx-vs-pos} (2 sources at $(\alpha,\delta)\sim(252^{\rm o},-14^{\rm o})$) the sample is reduced to 107 objects. The resulting weighted mean parallax for this reduced sample is $7.23\pm0.25$~mas.  
      
      To compare against the VLBA parallaxes, we bin the Gaia parallaxes into R.A.\ bins of width equal to 0.5$^{\rm o}$, and take the weighted average of the parallaxes within each bin. The resulting values are plotted in Figure  \ref{fig:oph-plx-vs-pos} as yellow triangles. The weighted parallaxes for the  two first bins are $6.82\pm0.40$ and $6.88\pm0.13$~mas, respectively, which agree within the errors with the VLBA parallax derived for L1689 ($6.93\pm0.06$~mas). Gaia data also suggest that the parallaxes at the center of L1688 are, on average, slightly larger than L1689. The weighted average parallax rises to $7.31\pm0.21$~mas  roughly at the center of L1688.

\begin{figure*}[!bht]
\begin{center}
 {\includegraphics[width=0.8\textwidth,angle=0]{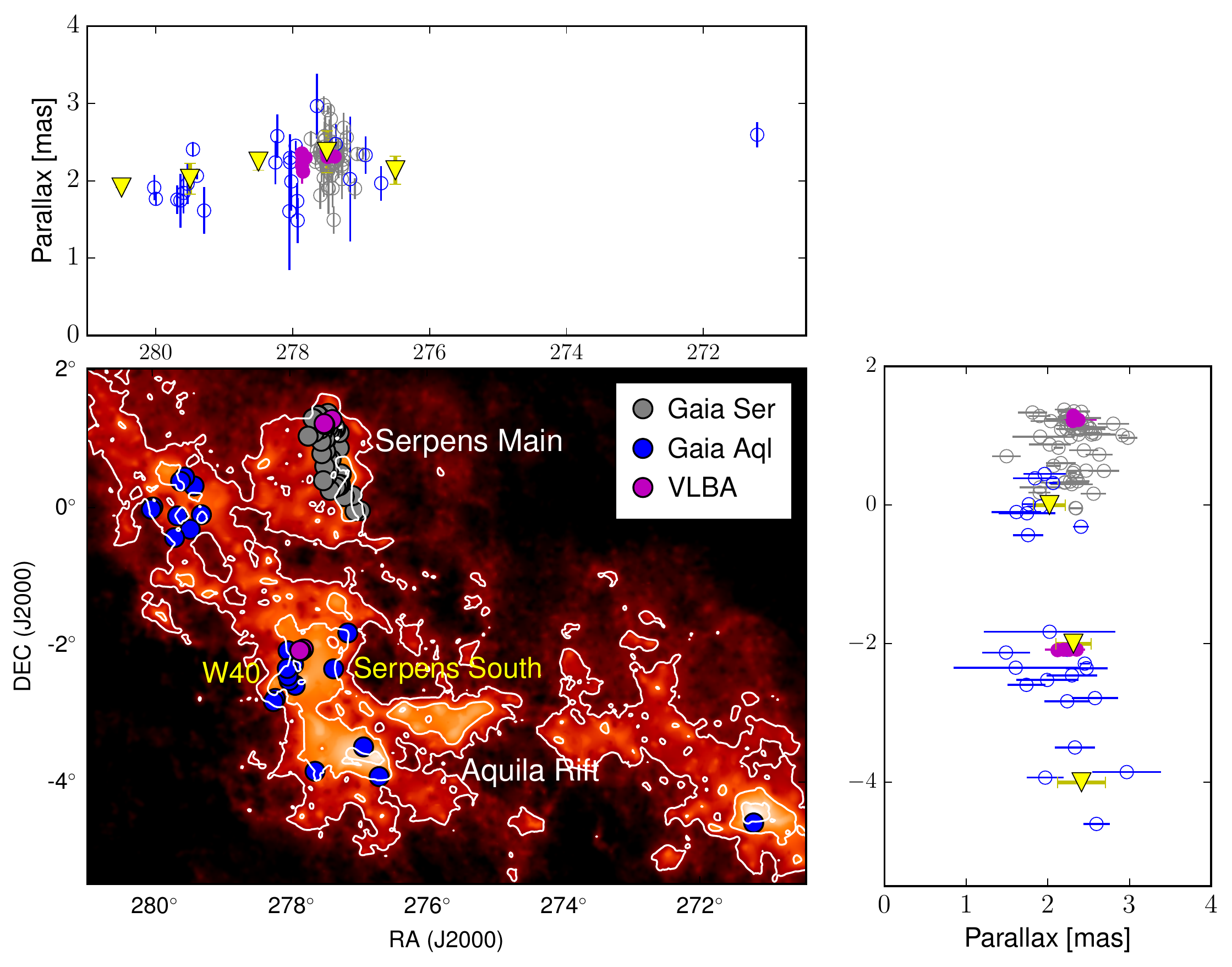}}
 \end{center}
\caption{Same as Figure \ref{fig:oph-plx-vs-pos} but for the Serpens/Aquila region. Here the bins have widths of $1^{\rm{o}}$ and $2^{\rm{o}}$, in R.A.\ and declination, respectively. The yellow triangles are the weighted mean of Gaia parallaxes in Aquila determined for these bins. Gaia parallaxes measured in Aquila are shown as blue circles. The extinction map in the background was taken from \cite{Cambresy1999}. The white contours indicate $A_V=$~4, 7 and 9. }
\label{fig:aquila}
\end{figure*}

For Aquila, we found 341 YSO candidates with available parallaxes in DR2. We can see in the middle panel of Figure \ref{fig:gaia-prlx} that the number of background contaminants is remarkably high for this region ($\sim 80\%$  of the total), and the presence of two distributions, with peaks separated at $\varpi\sim2$~mas, is clear. The stars with small parallaxes almost follow a Gaussian distribution as can be seen in the same figure. The center of this Gaussian is at $\varpi = 0.09\pm0.03$~mas and the width is $0.46$~mas. From this figure we can assume that all stars with $\varpi\apprle1.4$~mas, which is the boundary between the Gaussian and the second distribution,  are background contaminants. 
      We then cut further the sample by applying the proper motion criterium and the criteria to filter stars with spurious astrometric solutions as in Ophiuchus. This produced the reduced sample of 24 stars shown in Figure \ref{fig:aquila}, where we see that stars are grouped in at least three regions.  
      One group of stars is distributed within a diameter of $\sim$2 degrees around the position $(\alpha,\delta) \sim(277.6^{\rm o}, -2.3^{\rm o})$, encompassing the W40 and Serpens South clusters.  A second group of stars is located around the position $(\alpha,\delta) \sim(279^{\rm o}, +0^{\rm o})$. The next group is centred at $(\alpha,\delta) \sim(277^{\rm o}, -4^{\rm o})$  and finally  one more star is close to  $(\alpha,\delta) \sim(271^{\rm o}, -4.5^{\rm o})$. 

Similarly to Ophiuchus, we bin the Gaia parallaxes into R.A.\ and declination bins of width equal to $1^{\rm o}$ and $2^{\rm o}$, respectively. We then take the weighted average in each bin (the yellow triangles in Figure \ref{fig:aquila}).  The groups identified above in the spatial distribution of the Gaia parallaxes are clearly seen in the plot corresponding to the declination direction (right panel in Figure \ref{fig:aquila}). The weighted mean of stars within a radius of 1 degree around $(\alpha,\delta) \sim(277.6^{\rm o}, -2.3^{\rm o})$, i.e.\ the area covering the W40 and Serpens South clusters and corresponding to the second yellow triangle in the right panel of Figure \ref{fig:aquila}, gives $\varpi=2.31\pm 0.22$~mas. This is the representative Gaia value that we take for W40/Serpens South. 


In Serpens, only 68 {\it Spitzer} sources have parallaxes available in DR2. We also cross-matched the catalog of optical candidate young stars published by \cite{Erickson2015}, which contains 62 candidate members based on the presence of H$\alpha$ emission, lithium absorption, X-ray emission, mid-infrared excess, and/or reflection nebulosity.  For this catalog, we use a match radius of $2"$ to be more conservative. This adds 47 stars to the number of sources with Gaia parallaxes in Serpens. The parallax distribution is shown in the right panel of Figure \ref{fig:gaia-prlx}. Again, we see a peak in the parallax distribution due to background contaminants, which we remove by using the proper motion criterium. The weighted mean parallax for Serpens (of the 59 stars left after cleaning the sample from spurious astrometric solutions) gives $\varpi=2.32\pm 0.18$~mas. We take this value as the representative Gaia parallax for Serpens Main. 

\section{Discussion} 
 
The Gaia-DR2 measurements confirm the results reported in Papers I and II, which were based on VLBA data alone. In paper II, we derived  parallaxes for stars in W40 and Serpens Main. The VLBA parallaxes for both regions were consistent between them and suggested a common distance of $\sim 436$~pc. We argued that, given that Serpens South is projected very close to the W40 cluster, it should be located at the same distance of $\sim 436$~pc as W40, which implied that the three regions, i.e.\ Serpens Main, Serpens South and W40, are part of the same large  complex of molecular clouds. This conclusion was supported by the similar local standard of rest (LSR) velocities obtained from observations of isotopologues of CO toward IRAS~18275--0203 and a nearby embedded source in Serpens South \citep{Gutermuth_2008} and toward various positions within the Serpens Molecular cloud \citep{White_1995}.

The  values of the mean parallax derived from the Gaia-DR2 data alone for Serpens South/W40 and Serpens Main ($\varpi=2.31\pm 0.22$~mas and $\varpi=2.32\pm 0.18$~mas, respectively) are identical within the errors, which conclusively places both regions at a similar distance ($432^{+45}_{-37}$~pc and  $430^{+36}_{-30}$~pc, respectively), i.e. at a significantly larger distance than the old, but still used by some authors, value of 260~pc \citep{Straizys_2003}. As seen in the third panel of Figure \ref{fig:gaia-prlx}, there is a group of stars having smaller distances ($\varpi \apprge 4$~mas or $d\apprle 250$~pc), but they represent a minority compared to the main body of young stars, which draw a clear distribution around 2.3~mas.  

When compared against the VLBA parallaxes, we see an offset of $\varpi_{\rm Gaia} - \varpi_{\rm VLBA} = +0.09\pm0.07$~mas and $\varpi_{\rm Gaia} - \varpi_{\rm VLBA} = +0.02\pm0.02$~mas, for W40/Serpens South and Serpens Main respectively, where $\varpi_{\rm Gaia}$ and $\varpi_{\rm VLBA}$  are the weighted mean parallax for Gaia and the VLBA, respectively. 
      In Ophiuchus, we measure offsets of $\varpi_{\rm Gaia} - \varpi_{\rm VLBA}=+0.06\pm0.02$~mas and $\varpi_{\rm Gaia} - \varpi_{\rm VLBA}=-0.07\pm0.04$~mas, in L1688 and L1689, respectively. Although these offsets are relatively small, it is important to understand their origin. The offset measured for stars in L1689 is consistent with the zero-point error of Gaia DR2 parallaxes, which has a mean of $-0.029$~mas \citep[][]{Lindegren_2018}. The Gaia--VLBA parallax offset is positive in L1688, Serpens Main and the Aquila, which means that there the Gaia parallax is larger that the VLBA parallax, or that Gaia distances are, on average, smaller than the VLBA distances. This is actually the behavior we expect to see in  regions where the optical extinction is high ($A_V \apprge  20$ as is seen in Serpens and Ophiuchus; \citealt{Ridge_2006,Bontemps_2010}). Since Gaia observes in the optical, it can only detect stars that are not too obscured by dust extinction, while the VLBA can see through the extinction wall. Thus, Gaia could be biased toward detecting stars preferentially on the near of the clouds, whereas the VLBA does not suffer from such a bias. This effect is not evident in L1689 or IC~348 in Perseus \citep{Ortiz2018}, where the optical extinction is lower ($A_V \apprle 10$; \citealt{Ridge_2006} -- see also \citealt{Kounkel2018AJ} for a parallax comparison of the whole YSOs sample in common to Gaia and GOBELINS). 

      Then, our findings suggest Gaia would be biased in regions with strong optical extinction.  We were not  able to investigate the magnitude of the bias as a function of extinction from ancillary data. Visual extinction maps available in the literature have been derived from different datasets and using different algorithms. For instance, the Ophiuchus map shown in Figure \ref{fig:oph-plx-vs-pos} was obtained from 2MASS near-infrared data, while the Serpens map (Figure \ref{fig:aquila}) was derived from optical star counts and is only tracing the first layer of the extinction wall. From the distribution of the extinction measured in Ophiuchus, which increases from the east to the core of the cloud, we suggest that Gaia parallaxes on regions with $A_V~\apprge~18$ may show a significant bias. 
      

      It is also important to note that the error bars on Gaia parallaxes are on average significantly larger than the errors on VLBA parallaxes, mainly in the Aquila Region, and that the individual Gaia values show a larger dispersion than the dispersion seen for VLBA sources. This is clearly seen in Figures \ref{fig:oph-plx-vs-pos} and \ref{fig:aquila}. Quantitatively, this comes from taking the mean of Gaia parallax errors of the samples in each region before cleaning from possible poor astrometric solutions. The mean errors are $\sigma_{\varpi,{\rm Gaia}}=$~0.24, 0.18 and 0.40~mas in Ophiuchus, Serpens and Aquila, respectively, while the mean of VLBA parallax errors is $\sigma_{\varpi,{\rm VLBA}}=$~0.12~mas in the three regions. 
      
Given that the Gaia zero-point offset applicable to a given region in the sky is still not well determined \citep{Lindegren_2018}, and the corrections to be applied to the Gaia parallaxes are still poorly constrained, we do not attempt here to correct the Gaia parallaxes. We instead recommend to use the VLBA parallaxes and distances as they still represent the most accurate measurements obtained so far toward these three regions. Inverting the VLBA parallaxes yields distances of $138.4\pm2.6$~pc,   $144.2\pm1.3$~pc  and $436\pm9$~pc for L1688, L1689 and Serpens/Aquila, respectively.  
      
Regarding the presence of parallax gradients across the clouds, we do see  in Figures \ref{fig:oph-plx-vs-pos} and \ref{fig:aquila} that Gaia parallaxes change with position, rising toward the central part of the clouds, which is expected if extinction is biasing more toward the center. In Ophiuchus, VLBA parallaxes  show clear variations with position along the R.A.\ direction, which supports  the idea that different structures in the cloud are located at slightly different distances.  In Serpens/Aquila, the significance of such parallax variations is low. However, we do not rule out a possible difference of $\sim 16$~pc between Serpens Main and Serpens South/Aquila. 
      
\begin{figure*}[!bht]
\begin{center}
  {\includegraphics[width=0.48\textwidth,angle=0]{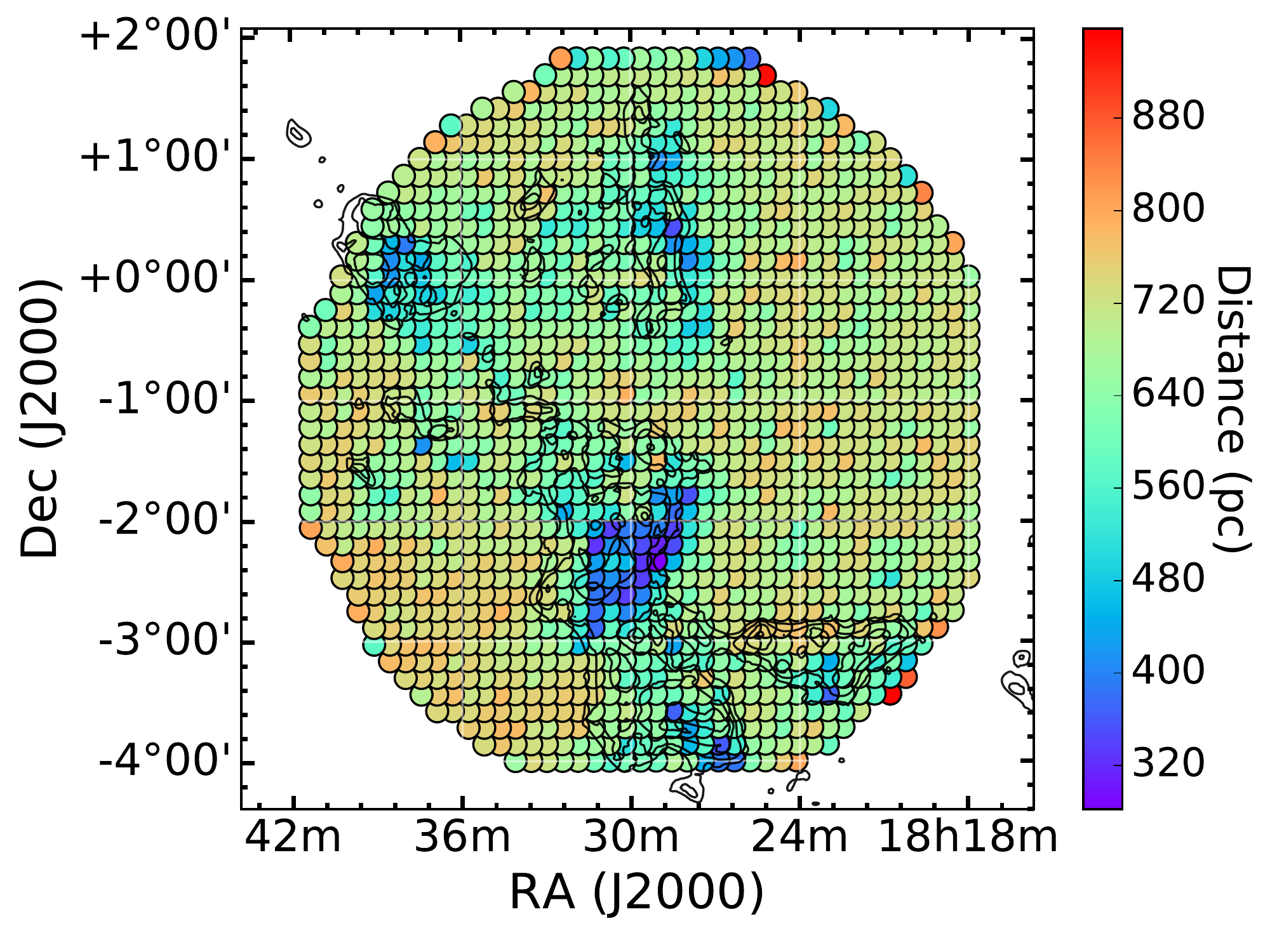}
 {\includegraphics[width=0.45\textwidth,angle=0]{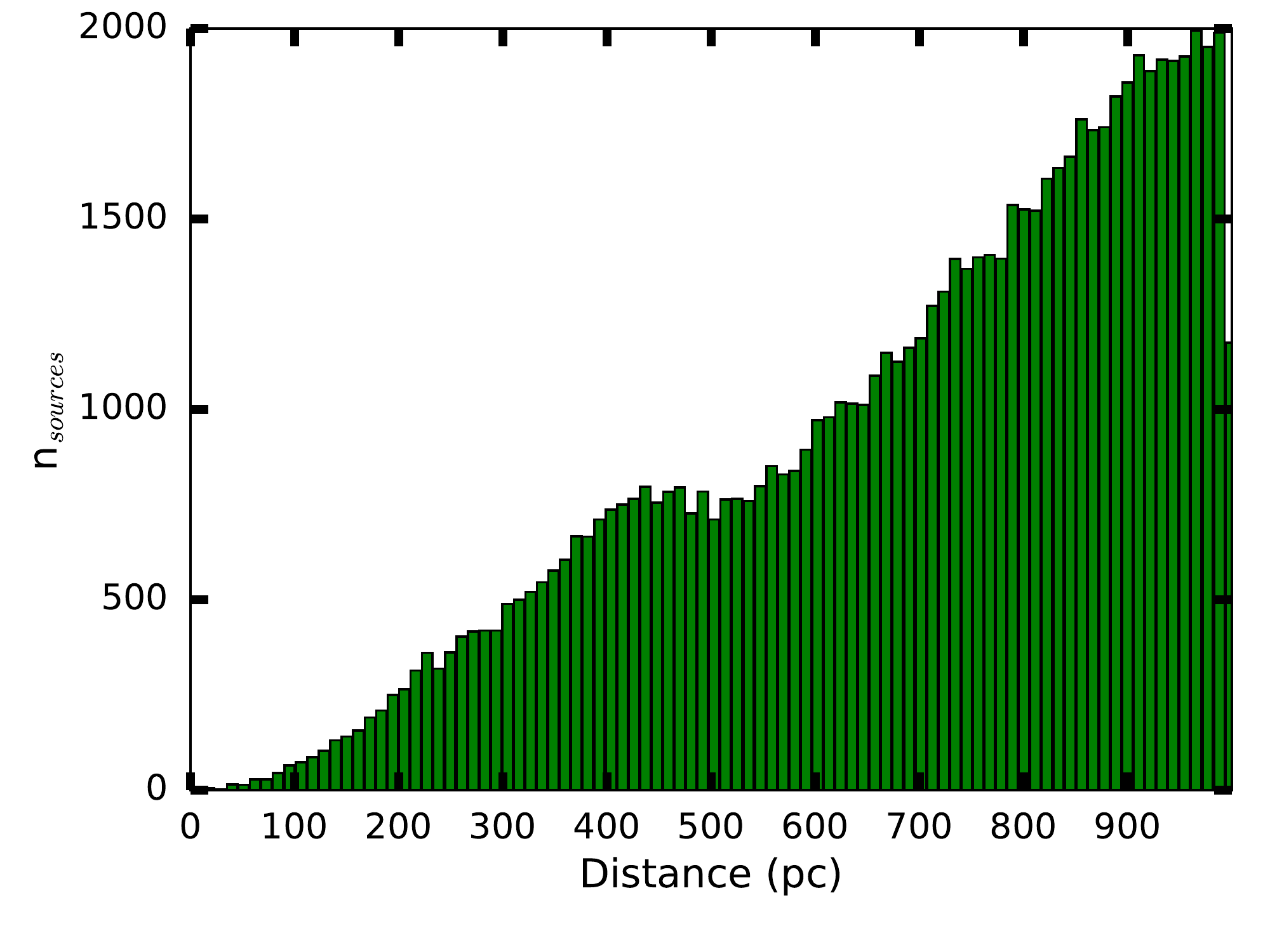}}}
 \end{center}
\caption{ 
{\it Left:} Gaia-DR2 distances for all stars within $3^{\rm o}$ around $(\alpha,\delta) = (277.5, -1.0^{\circ})$ and with $\varpi>1$~mas and  $\sigma_\varpi<0.5$~mas, overlaid on the extinction map of the Aquila.  Each point is the average of all stars within small patches of size set to 500~arcsec. {\it Right:} Histogram of  distances for the same sample of stars. }
\label{fig:mapa}
\end{figure*}

To  strengthen our conclusion on the distance to Aquila, we use all stars from the Gaia DR2 catalog covering a circular area with  a radius of 3$^\circ$ and centered at  $(\alpha,\delta) = (277.5, -1.0^{\circ})$ (i.e., at the mid-point between Serpens Main and Serpens South). We only keep stars with $\varpi>1$~mas ($d<1$~kpc) and $\sigma_\varpi<0.5$~mas, resulting in a total of 79550 stars. The spatial distribution of these stars is shown in Figure~\ref{fig:mapa}, where each point represent the average of all stars within small patches of size set to 500~arcsec, whose average distance (estimated as $1/\overline \varpi$) is indicated by a color code.  We see in Figure~\ref{fig:mapa} the effect of the obscuration by the cloud, because, on average, the areas with  large extinction are the ones that have the lower distances. The distribution of distances is shown also in Figure~\ref{fig:mapa}. Here we see that the number of sources keeps rising with distance until  it reaches $\sim 430$~pc, where there is a pause in the rise. This distance  corresponds to the main distance to the cloud, which agrees very well with the distance derived from the VLBA parallax measurements. 

The nature of the background {\it Spitzer} contaminants in Aquila is still unknown, but they are likely AGB stars according to \cite{Dunham_2015}. The proper motions of these stars are non-zero, which suggests they are Galactic sources. 

\section{Conclusions}
We used Gaia-DR2 and VLBA data to investigate the distance to Ophiuchus, Serpens Main and W40/Serpens South in the Aquila Rift region. Our target samples for the Gaia analysis consisted of YSO candidates with infrared excess  identified  by {\it Spitzer}, which were complemented by including 2MASS and optical sources with other signs of an association to young stars. We carefully cleaned our samples from background and foreground stars, and from stars with spurious astrometric solutions, and then derived the weighted mean parallax of each region. From Gaia data alone, we found identical parallaxes for both Serpens Main ($\varpi_{\rm Gaia}=2.32\pm 0.18$~mas) and W40/Serpens South ($\varpi_{\rm Gaia}=2.31\pm 0.22$~mas), which are consistent with the parallaxes measured independently by the VLBA ($\varpi_{\rm VLBA}= 2.30 \pm 0.05$~mas). Hence, this confirms that the three regions are part of the same complex of molecular clouds. VLBA and Gaia-DR2 parallaxes are also highly consistent between them in Ophiuchus, where we found $\varpi_{\rm L1688, VLBA}=7.23\pm 0.14$~mas,  $\varpi_{\rm L1689, VLBA}=6.93\pm 0.06$~mas, $\varpi_{\rm L1688, Gaia}=7.29\pm 0.22$~mas and $\varpi_{\rm L1689, Gaia}=6.86\pm 0.23$~mas. The small offsets between Gaia and the VLBA can be understood in terms of the Gaia parallax zero point error or Gaia being biased toward the brightest and less obscured stars.

\acknowledgements{G.N.O.-L. acknowledges support from the von Humboldt Stiftung. L.L. acknowledges the financial support of DGAPA, UNAM (project IN112417), and CONACyT, M\'exico.

The Long Baseline Observatory is a facility of the National Science Foundation operated under cooperative agreement by Associated Universities, Inc. The National Radio Astronomy Observatory is a facility of the National Science Foundation operated under cooperative agreement by Associated Universities, Inc.  

This work has made use of data from the European Space Agency (ESA) mission {\it Gaia} (\url{https://www.cosmos.esa.int/gaia}), processed by the {\it Gaia} Data Processing and Analysis Consortium (DPAC, \url{https://www.cosmos.esa.int/web/gaia/dpac/consortium}). Funding for the DPAC has been provided by national institutions, in particular the institutions participating in the {\it Gaia} Multilateral Agreement.
}

\clearpage
\begin{deluxetable}{cccccccc}
\tabletypesize{\scriptsize}
\tablewidth{0pt}
\tablecolumns{8}
\tablecaption{Astrometric parameters from VLBA measurements \label{tab:prlx-oph} }
\tablehead{ GBS-VLA& Other identifier & Parallax & $\mu_\alpha\cos\delta$ & $\mu_\delta$      & a$_\alpha\cos\delta$ & a$_\delta$         &Distance \\
                   name      &                         & (mas)     & (mas yr$^{-1}$)               & (mas yr$^{-1}$)  & (mas yr$^{-2}$)          & (mas yr$^{-2}$) & (pc)          \\
                   (1)           &           (2)         &    (3)       &   (4)                                 &   (5)                    &   (6)                            &  (7)                    & (8)            \\
}
\startdata
\cutinhead{Ophiuchus} %
J162556.09-243015.3 & WLY2-11&  7.181 $\pm$ 0.135 & -8.098 $\pm$ 0.075 & -21.978 $\pm$ 0.222 & 1.828 $\pm$ 0.14 & 4.264 $\pm$ 0.412 &  139.26 $\pm$ 2.62  \\ 
J162557.51-243032.1 & YLW24 &  7.089 $\pm$ 0.458  & -7.047 $\pm$ 0.198 & -25.135 $\pm$ 0.255 & -- & --  & 141.07 $\pm$ 9.12 \\
J162603.01-242336.4 & DoAr21 & 7.366 $\pm$ 0.123 & -19.567 $\pm$ 0.064 & -26.936 $\pm$ 0.056 & -- & -- & 135.76 $\pm$ 2.27 \\ 
J162622.38-242253.3 & LFAM2 & 7.216 $\pm$ 0.119 & -5.622 $\pm$ 0.125 & -26.658 $\pm$ 0.308 & -- & --  &  138.59 $\pm$ 2.28 \\  
J162629.67-241905.8 & LFAM8 & 7.254 $\pm$ 0.112 & -5.983 $\pm$ 0.059 & -29.642 $\pm$ 0.129 & -- & -- & 137.86 $\pm$ 2.13  \\ 
J162634.17-242328.4 & S1        & 7.364 $\pm$ 0.052 &  -2.070 $\pm$ 0.010 & -26.790  $\pm$  0.021 & -- & -- & 135.79 $\pm$ 0.97 \\  
J162642.44-242626.1 & LFAM15 & 7.176 $\pm$ 0.058 & -6.327   $\pm$  0.016 & -26.905  $\pm$  0.036 & -- & -- & 139.36 $\pm$ 1.13    \\ 
J162643.76-241633.4 & VSSG11 & 7.160  $\pm$ 0.152 & -10.48 $\pm$ 0.16 & -38.99 $\pm$ 0.35 & 0.31 $\pm$ 0.65 & -1.54 $\pm$ 1.09 & 139.7 $\pm$ 3.0 \\
J162649.23-242003.3 & LFAM18 & 7.176 $\pm$ 0.055 & -8.627 $\pm$ 0.159 &  -20.015 $\pm$ 1.064 & -- & -- & 139.35 $\pm$ 1.06   \\ 
J162718.17-242852.9 & YLW 12Bab & 7.135 $\pm$ 0.064 & 10.875 $\pm$ 0.049 & -25.057 $\pm$ 0.140 &  -0.622 $\pm$ 0.116 &  -0.177 $\pm$   0.018 & 140.16 $\pm$ 1.26  \\   
J162718.17-242852.9 & YLW 12Bc$^1$ & 7.135 $\pm$ 0.064 & -11.219 $\pm$ 0.052 & -23.107 $\pm$ 0.128 & 0.195 $\pm$ 0.027 & 0.109 $\pm$ 0.066 & 140.16 $\pm$ 1.26 \\ 
J162721.81-244335.9 & ROXN39 & 7.396 $\pm$ 0.071  & -6.858 $\pm$ 0.095 & -24.864 $\pm$ 0.146 & -- & --  & 135.22 $\pm$ 1.30  \\ 
J162721.97-242940.0 & GY256  & 6.711 $\pm$ 0.111 & -6.673 $\pm$ 0.054 & -34.41 $\pm$ 0.176 &   -- & --   & 149.02 $\pm$ 2.47 \\
J162726.90-244050.8 & YLW15  & 7.376 $\pm$ 0.095 & -12.841 $\pm$ 0.042 & -26.255 $\pm$ 0.169 &   -- & --   & 135.57 $\pm$ 1.75 \\
J162730.82-244727.2 & DROXO71 &  7.455 $\pm$ 0.229 & -4.822 $\pm$ 0.144 & -28.256 $\pm$ 0.253 & -- & -- & 134.14 $\pm$ 4.12 \\ 
J162804.65-243456.6 & ROXN 78 & 7.185 $\pm$ 0.091 & -5.446 $\pm$ 0.028 & -29.301 $\pm$ 0.071 & -- & -- & 139.18 $\pm$ 1.77   \\ 
J163035.63-243418.9 & SFAM 87 &  7.216 $\pm$ 0.068  &  -7.702 $\pm$ 0.019 & -26.028 $\pm$ 0.031 & -- & -- &  138.58 $\pm$ 1.31  \\ 
J163115.01-243243.9  & ROX42B & 6.922 $\pm$ 0.043 & -5.817 $\pm$ 0.033 & -23.206 $\pm$ 0.259 & -0.506 $\pm$ 0.041 & 0.962 $\pm$ 0.315 & 144.47 $\pm$ 0.91 \\ 
J163151.93-245617.4 &  --  &  7.265  $\pm$ 0.778   &   -8.564 $\pm$ 0.268 & -27.132 $\pm$ 0.589 & -- & -- &  $138^{+17}_{-13}$ \\
J163152.10-245615.7 & LDN1689IRS5 & 6.677 $\pm$ 0.157 & -6.537  $\pm$ 0.09 &  -22.557 $\pm$  0.144 &  -- & --  & 149.76  $\pm$ 3.52 \\ 
J163200.97-245643.3 & WLY2-67 & 6.741 $\pm$ 0.173 & -5.699 $\pm$ 0.14 & -23.994 $\pm$ 0.357 & -- & --  &  148.34 $\pm$ 3.80 \\ 
J163211.79-244021.8 & DoAr51 &   6.972 $\pm$ 0.041 & -4.746  $\pm$  0.084 &  -23.139 $\pm$ 0.099 &  -- & -- &  143.43 $\pm$  0.85  \\ 
\cutinhead{Serpens Main} %
J182933.07+011716.3  & GFM 11  & 2.313 $\pm$ 0.078  & 3.634  $\pm$ 0.050  & -8.864 $\pm$ 0.127  &  -- & -- & $432^{+15}_{-14}$  \\
J182957.89+011246.0  &  EC 95    & 2.307 $\pm$ 0.022 & 3.579 $\pm$ 0.021  &  -8.359 $\pm$  0.023  &   --  & -- & 433 $\pm$ 4 \\
J183000.65+011340.0  & GFM 65  & 2.375 $\pm$ 0.222 & 2.437 $\pm$ 0.357  & -8.263 $\pm$ 0.228  & -- & -- &  $421^{+44}_{-36}$  \\
\cutinhead{W40} %
J183114.82$-$020350.1 & KGF 36 &  2.297 $\pm$ 0.116  & 0.400 $\pm$ 0.065 & -6.607 $\pm$ 0.063 &  --  &  --  &  $435^{+23}_{-21}$ \\ 
J183123.62$-$020535.8 & KGF 97 &  2.119  $\pm$ 0.077 & -0.300 $\pm$ 0.047 & -7.432 $\pm$ 0.048  &  --  &  --  &  $472^{+18}_{-17}$ \\
J183126.02$-$020517.0 & KGF 122 & 2.261 $\pm$ 0.138 & 1.845 $\pm$ 0.645  & -6.029 $\pm$ 0.339   &  --  &  --  &  $442^{+29}_{-25}$  \\  
J183127.45$-$020512.0 & KGF 133 & 2.194 $\pm$ 0.231 & 0.181 $\pm$ 0.148  & -8.750 $\pm$ 0.323   & 0.270 $\pm$ 0.158 & -0.53 $\pm$ 0.344 & $456^{+54}_{-44}$  \\ 
J183127.65$-$020509.7 & KGF 138&  2.353 $\pm$ 0.106 & 0.172 $\pm$ 0.089 & -6.784 $\pm$  0.785  &  --  &  --  &  $425^{+20}_{-18}$ \\ 
%
%
\enddata
\tablenotetext{1}{Parallax is fixed at the value obtained for YLW 12Bab when solving for the other astrometric parameters. 
}
\end{deluxetable}
\clearpage


\end{document}